\documentclass[10pt]{article}

\usepackage{amsmath}
\usepackage{amssymb}

\usepackage{graphicx}
\usepackage{epstopdf}

\usepackage{cite}
\usepackage{color} 

\usepackage{setspace} 

\usepackage[labelfont=bf,labelsep=period,justification=raggedright]{caption}

\makeatletter
\renewcommand{\@biblabel}[1]{\quad#1.}
\makeatother

\date{}
\begin{document}

\begin{flushleft}
{\Large
\textbf{A fast and scalable kymograph alignment algorithm for nanochannel-based optical DNA mappings}
}
\\
Charleston Noble$^{1, 2, 4}$,  
Adam N. Nilsson$^{1}$, 
Camilla Freitag$^{2, 3}$,
Jason P. Beech$^{2}$,
Jonas O. Tegenfeldt$^{2}$,
Tobias Ambj{\"o}rnsson$^{1, \ast}$
\\
\bf{1} Department of Astronomy and Theoretical Physics, Lund University, Lund, Sweden
\\ 
\bf{2} Division of Solid State Physics, Department of Physics, Lund University, Lund, Sweden
\\ 
\bf{3} Department of Physics, Gothenburg University, Gothenburg, Sweden
\\
\bf{4} Current address: Department of Systems Biology, Harvard Medical School, Boston, MA 02115, USA
\\
$\ast$ E-mail: tobias.ambjornsson@thep.lu.se
\end{flushleft}


\section*{Abstract}
Optical mapping by direct visualization of individual DNA molecules, stretched
in nanochannels with sequence-specific fluorescent labeling, represents a
promising tool for disease diagnostics and genomics. An important challenge
for this technique is thermal motion of the DNA as it undergoes imaging; this
blurs fluorescent patterns along the DNA and results in information
loss. Correcting for this effect (a process referred to as kymograph
alignment) is a common preprocessing step in nanochannel-based optical mapping
workflows, and we present here a highly efficient algorithm to accomplish this
via pattern recognition. We compare our method with the one previous approach, and we find that our method is orders of magnitude faster while producing data of similar quality. We demonstrate proof of principle of our approach on experimental data consisting of melt mapped bacteriophage DNA.

\section*{Introduction}
Optical mapping is an emergent complementary approach to DNA sequencing which produces a lower resolution (typically kbp) sequence-dependent map of individual DNA molecules\cite{opticalmapping1,opticalmapping2, neely2011optical, das2010single, xiao2007rapid, valouev2006algorithm, valouev2006alignment, ananiev2008optical, jing1998automated, cai1998high}. As a direct complement to DNA sequencing, optical mapping can produce a scaffold to facilitate easier sequence assembly, and as an indirect complement, optical mapping promises a variety of applications for which low resolution maps will suffice. For example, it can be used to quickly identify large-scale structural variations, including duplications, deletions, insertions, inversions, and translocations, which are increasingly being linked to heritable traits of phenotypic significance \cite{variationProblems1,variationProblems2, neely2011optical}, and it allows for the rapid identification of bacterial species and strains which could represent an important step against the growing problem of antibiotic resistance \cite{bacteria1,bacteria2,bacteria3,bacteria4,bacteria5}.

To date, many techniques for optical mapping have been developed, and they typically rely on sequence-specific DNA modifications at short target sites, followed by imaging and analysis. These sequence-specific modifications can include staining and denaturation ({}``melt mapping'') \cite{reisner2010single}, fluorocoding \cite{fluorocode}, competitive binding of intercalating dyes \cite{competitiveBinding, nilsson2014competitive}, methylation \cite{methylation}, enzymatic nicking \cite{enzNicking1,enzNicking2}, and enzymatic restriction \cite{restriction1,restriction2,restriction3}.

These optical mapping techniques can roughly be divided into  three groups:
stretching over a surface\cite{jing1998automated, schwartz1993ordered,
  cai1995ordered, bensimon1994alignment}, stretching via confinement in
nanochannels\cite{jo2007single, douville2008dna, das2010single,
  reisner2005statics, reisner2012dna, tegenfeldt2004dynamics, persson2010dna}
 or stretching via elongational flow in
  microchannels\cite{marie2013integrated}. While surface--stretching
  techniques offer a few advantages, such as allowing for 100-140\%
  extension\cite{restriction2}, mapping via nanochannel confinement allows for
  integration of the stretching in a lab on a chip context that in turn can be
  brought to application much more easily. 

Note that the nanochannel-based DNA barcoding schemes, the main focus of this study, should not be confused with schemes where short genetic markers\cite{hebert_2003} or restriction enzyme cleavage events\cite{restriction1} are detected as landmarks. Barcodes addressed by our technique are not simply binary in the sense that one detects landmarks or the absence thereof; rather what we call ``barcodes" are continuous fluorescence profiles which are more susceptible to thermal noise (see Fig. \ref{fig:problemDefinition}a).

One particular problem inherent to nanochannel confinement techniques is that DNA tends to undergo random diffusive processes during imaging, including center-of-mass diffusion and local stretching\cite{diffusion}. To correct for these effects, a procedure we denote \emph{kymograph alignment} must be performed (see section Problem Definition). In this paper we present WPAlign (\textbf{W}eighted \textbf{P}ath
\textbf{Align}), an algorithm for kymograph alignment which offers
linear scaling in time and can align DNA barcodes with length corresponding to
an entire human genome in less than an hour on a typical desktop computer. We compare its
performance to an existing technique and show that our method offers orders-of-magnitude
improvement in computational speed while producing processed data of similar quality.
Additionally, we present a new information score which
quantifies the information content of DNA barcodes and should see widespread 
use as a barcode quality criterion by which experimentalists can evaluate barcodes
and optimize experimental mapping conditions.

\section*{Problem Definition}
\begin{figure}
\noindent \begin{centering}
\includegraphics[width=0.75\textwidth]{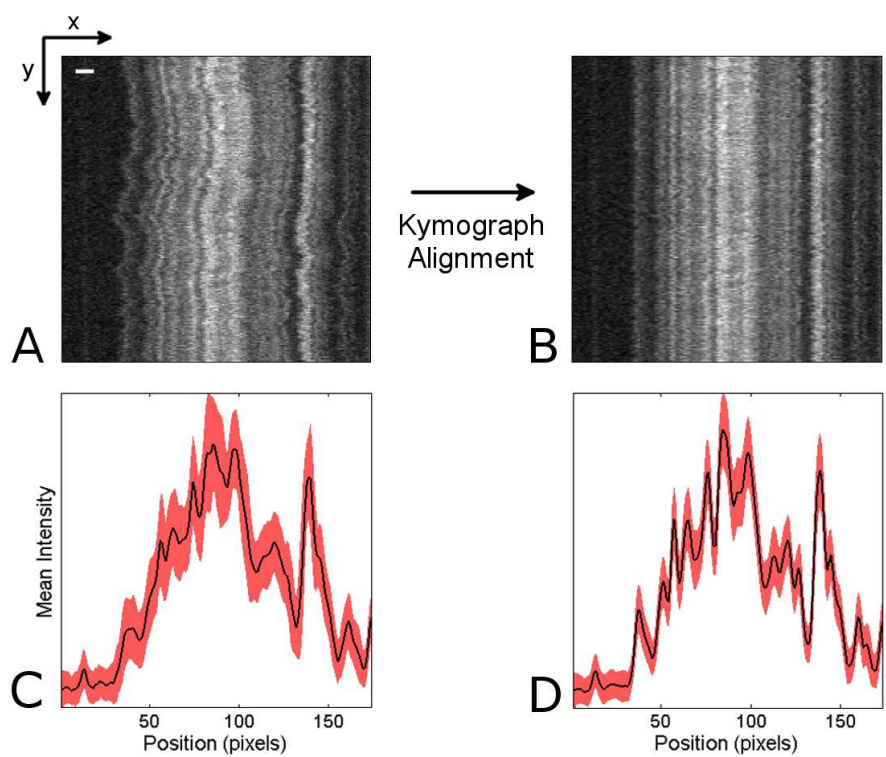}
\par\end{centering}

\caption{{\bf Problem definition.} (A) A raw kymograph depicting denaturation mapping of an intact T4GT7 bacteriophage DNA molecule. The y-axis represents time (200 time frames over 20 seconds) while the x-axis represents the DNA extension in the nanochannel (170 pixels wide; scale bar is 2$\mu$m). Note that these specifications also apply to all of the following kymographs. (B) The image from (A) after undergoing the kymograph alignment process. (C) Trace of mean intensity over time for the raw kymograph (black) with $\pm1$ standard deviation (red), and (D) the same for the aligned kymograph. 
\label{fig:problemDefinition}}
\end{figure}

In Figure \ref{fig:problemDefinition}a, the result of a typical nanochannel-based optical DNA
mapping experiment is displayed. The horizontal axis (x-axis) of this
kymograph represents the DNA's extension
in the nanochannel (where the pattern results from sequence-specific staining), 
and the vertical (y-axis) represents images of the
DNA molecule at different times. 

Due to thermal center of mass
and local conformational fluctuations (which occur because the flexible DNA
molecule is only
extended to roughly 50 percent of its contour length), the images over time
are misaligned both locally and at a global
level, leading to significant noise during time-averaging (see Figure \ref{fig:problemDefinition}c).
Thus before a useful time-average can be performed, the bright
and dark bands must be ``straightened"; this procedure we refer to as \emph{kymograph alignment}. 
Figure \ref{fig:problemDefinition}b shows the result of
our algorithm, WPAlign, developed for this particular task, and
Figure \ref{fig:problemDefinition}d shows the corresponding time-average.
The details of
the algorithm can be found in the Methods section. Intuitively, the purpose of
this process is to, as closely as possible, mimic the results which would be obtained
if the DNA were held and stretched to 100 percent of its length during imaging.
For practical experimental reasons, this is not feasible for nanochannel-based
techniques, so one must correct for misalignments such as those displayed in Fig. \ref{fig:problemDefinition}a.

 In Ref. \cite{reisner2010single}, the kymograph alignment challenge is
 mathematically presented as a global optimization problem (see
 S1 for details): a template time frame, $T$, is first
 chosen. One then allows for local stretching of all the remaining
 frames, $N_i$, and maximizes the overlap between $N_i$ and
 $T$. Finally, all the time frames are rescaled with average global
 stretching factors in order to ensure that the end-to-end
 distance of the DNA molecule after alignment is unchanged and that
 the time-averaged optical map is therefore consistent with
 thermodynamic constraints.

In this study we pose the kymograph alignment challenge in slightly
different mathematical terms, namely as a feature detection problem
(details found in Methods). Visually, our method (see Results)
produces appealing images which are
very similar to those produced using the previous method. The main
benefit of our approach is that it provides an orders-of-magnitude
reduction in computational time. Furthermore, all the substeps in our
algorithm use standard numerical tools and are, therefore, straightforward to
implement.

The rest of this study is organized as follows: in Methods we
introduce our kymograph alignment algorithm, WPAlign. In Results, we compare
our new algorithm to the previous method. In Discussion we summarize
our results and point to further potential applications of our
algorithm, and in the S1 we provide more technical details of
our approach. 

\noindent %

\section*{Methods}

We now consider the problem defined in the previous section, i.e., the alignment of a ``fuzzy" DNA barcode (see Fig. \ref{fig:problemDefinition}a for an illustrative example) distorted by thermal fluctuations as it resides in a nanochannel. WPAlign works  intuitively by detecting pronounced bright or dark ridges (``features'') and then stretching the barcode horizontally to straighten them. The advantage of this approach is that it reduces the problem to a simple two-step process of (1) {\it single-feature detection}, followed by (2) {\it single feature alignment}. This two-step process is then be applied recursively to align the entire kymograph.

\subsection*{Single Feature Detection}

Consider an image represented as a 2D gray scale intensity function with
columns $x$ and rows $y$, denoted $I(x,y)$, $x\in[1,n],$ $y\in[1,m]$. See
Fig. \ref{fig:problemDefinition}a for an illustrative example.  Then a feature
$x=F(y)$ is a function mapping of rows $Y$ to columns $X$.  This
  mapping thus provides a  curve, $(x,y)=(F(y),y)$, through the image.
Intuitively the feature $F$ we would like to detect is the ``most pronounced''
vertical ridge or valley in $I$ such that:
\begin{enumerate}
\item (Completeness) Our detected feature $F$ traverses each row 
$y$ exactly once. Actual features might not adhere to this 
constraint, since anomalous events such as DNA breakage could occur, but 
these events should be treated separately from the alignment process.
\item (Continuity) $F$'s horizontal movement is constrained. 
I.e., $\lvert F(y+1)-F(y)\rvert\leq k$
for some integer $k$ and for all $y$. This is to fit the physical
constraints we inherit from the polymer physics of DNA molecules for
the present application.
\end{enumerate}
Unfortunately, standard feature detection (e.g., ridge detection)
methods cannot be applied directly, as the resulting features will
not necessarily satisfy these constraints, so we have developed
the following three-step procedure for {\it single feature detection}:

\subsubsection*{1. Image pre-processing}

\begin{figure}
\noindent \begin{centering}
\includegraphics[width=0.9\textwidth]{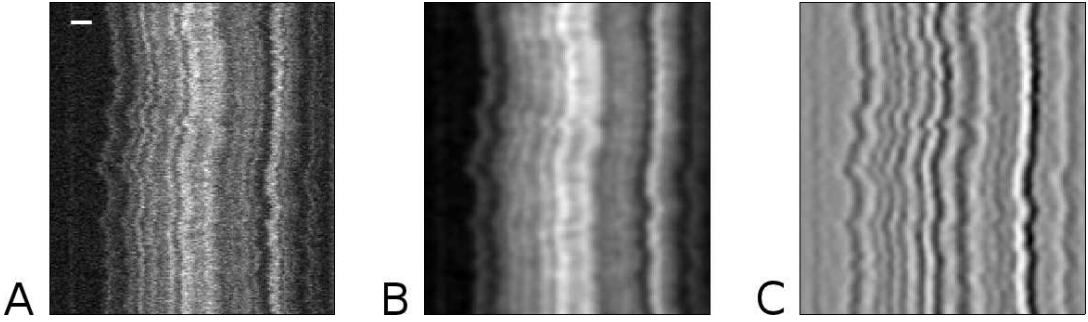}
\par\end{centering}

\caption{{\bf Image preprocessing.} (A) The raw, unaligned, kymograph shown in Fig. \ref{fig:problemDefinition}a (scale bar is 2 $\mu$m, 20 seconds of imaging).
(B) The kymograph after Gaussian
smoothing, and (C) The Laplacian response image, which we refer to as $K$, that results from applying
the Laplacian of Gaussian filter (with $\sigma = 10$ pixels) to our smoothed kymograph. In $K$, dark regions represent
positive values and light regions represent negative values.
\label{fig:PreProcessing}}
\end{figure}

First the entire image is smoothed with a 2D Gaussian kernel (here we use
$\sigma_{h}=10$ pixels and $\sigma_{v}=3$ pixels, where $h$ and $v$ represent
the horizontal and vertical directions, respectively) to remove noise from
random intensity fluctuations (see Fig.  \ref{fig:PreProcessing}b and
Fig. \ref{alg:WPAlign}, line 13). Then we apply a Laplacian of Gaussian filter
to obtain the {\it Laplacian response} image which we refer to as $K$ (see
Fig. \ref{fig:PreProcessing}c, and Fig. \ref{alg:WPAlign}, line 14), a matrix
with the same size as $I$ which has large positive values in dark bands and
large negative values in light bands (see also S1). These positive
and negative values are linearly rescaled to range from $(0,1]$ and $[-1,0)$,
respectively.

Now our feature $F$ should be represented by a continuous region of high
(valley) or low (ridge) values in $K$. We would like to treat these as two
distinct cases to avoid detecting a single feature which is partially composed
of a ridge and partially composed of a valley, so we compute two separate
images, $K_{B}$ and $K_{D}$ which emphasize bright and dark regions,
respectively (see Fig. \ref{fig:kDandkBandNetwork}, Fig. \ref{alg:WPAlign}
line 15, and S1). Both are positive or zero everywhere, with the
smallest values representing the most pronounced feature locations.

\noindent %

\subsubsection*{2. Network assembly}

\begin{figure}
\noindent \begin{centering}
\includegraphics[width=0.6\textwidth]{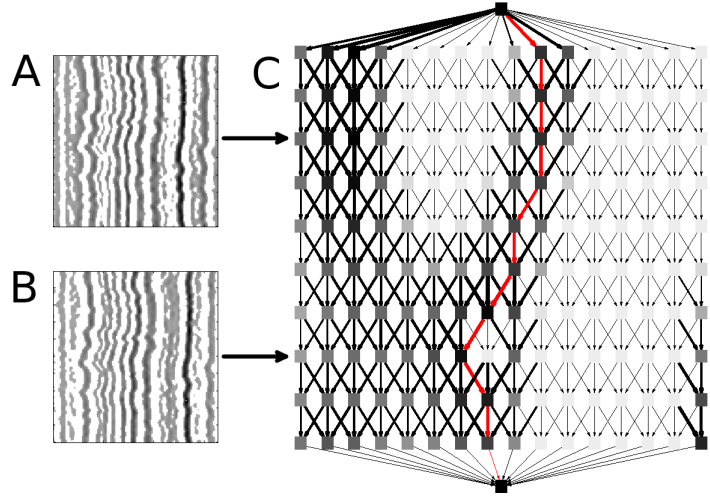}
\par\end{centering}

\caption{{\bf Network assembly.} The Laplacian response image, $K$, (see Fig. \ref{fig:PreProcessing}c) 
has been rescaled into (A)
$K_{D}$ and (B) $K_{B}$, images that emphasize dark and bright regions in $K$, respectively.
White pixels represent barriers that potential
features cannot cross, while continuous dark regions indicate likely features. 
(C) Here we show one example network, although separate networks are indeed assembled for 
$K_B$ and $K_D$. Each node (square) within the rectangular
region represents a pixel in K$_{B}$ or $K_{D}$. The top and bottom
nodes (which we term ``peripheral nodes") are added to provide starting/ending points for the
shortest path finding algorithm. The width of the edges corresponds to the
inverse of the edge weight, and the darkness of the nodes represents the average
weight of incoming edges. The red line illustrates the shortest path
through the network. For the sake of visual clarity, this network was 
created using a small subsection
of an actual Laplacian response with $k=1$.\label{fig:kDandkBandNetwork}}
\end{figure}

We can think of the images $K_{B}$ and $K_{D}$ as energy landscapes,
and finding the best ridge or valley becomes a problem of finding
the lowest-energy paths through $K_{B}$ and $K_{D}$, respectively.
To do this, we assemble directed, acyclic networks $G_{B}$ and $G_{D}$
as follows (see Fig. \ref{fig:kDandkBandNetwork}, and Fig. \ref{alg:WPAlign},
line 18). For simplicity, we describe our process only for $G_B$, as the
process is identical for $G_D$.

\begin{enumerate}
\item [1.]First, we create one node for every pixel in $K_{B}$, plus two
``peripheral" nodes. Thus $G_{B}$ consists of $(m\times n+2)$ nodes.
\item [2.]Now the first of the peripheral nodes is connected to each of
the nodes corresponding to the first row of $G_B$, and each of the
last-row nodes is connected to the second peripheral node with edge
weight 1. 
\item [3.]The rest of the nodes are connected as follows: the node corresponding
to pixel $(i,j)$ in $K_{B}$ is connected to pixels in the next row
directly below, to the left $k$ columns, and to the right $k$ columns
 (see Fig. \ref{fig:Tree}). This $k$ value is used to satisfy
the continuity constraint above. We have found $k=2$ to be reasonable for our
current application.
\begin{figure}
\noindent \begin{centering}
\includegraphics[width=0.7\textwidth]{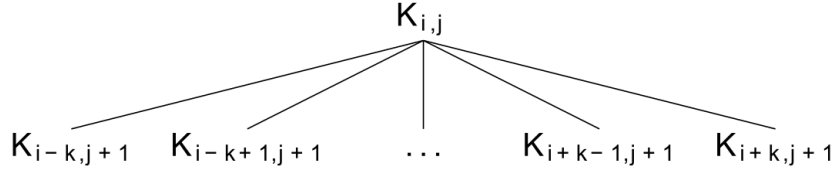}
\caption{{\bf Connectivity between nodes.}  
\label{fig:Tree}}
\par\end{centering}
\end{figure}

\item [4.] If a pixel is too close to a border on the left or right (i.e.,
$i-k<1$ or $i+k>n$), connections to non-existing nodes prescribed
in step 3 are ignored.
\item [5.] Finally, every edge is assigned a weight equal to the intensity
of the pixel in $K_{B}$ corresponding to the node it is directed to.
\end{enumerate}

Note that the edge weights in this connection scheme are not ``biased," in the 
sense that given equal intensities, movement to any of the connected pixels is
equally likely. Alternatively, one could assign a non-uniform distribution such
that the edge weights are given by a combination of intensities and some prior
knowledge of likely fluctuations. For example, the weight of the edge connecting
$(i,j)$ to $(i',j')$ could be given by $w_{i\to i', j\to j'} = I(i', j')f(i',i)$,
where $f$ is some probability distribution.
Our scheme here represents a special uniform case where $f(i',i) = 1 / (2k + 1)$ for 
$\mid i' - i\mid \leq k$ and $f(i',i)=0$ otherwise.

After examining several additional forms for $f$, we found that the simple uniform
case presented above achieved the best and most consistent results for our examples.
However, this weighting scheme should, in general, be carefully chosen by the 
user to suit the particular data being considered.

As for the choice of $k$, we found $k=2$ to be reasonable for the data
presented here; now we present a method for choosing $k$ to be used on novel
data. For unbiased diffusive motion, the diffusion length $\Delta L$ of the
DNA molecule is connected to time as $\Delta L^2 \propto Dt$ (where $t$ is the
inverse sampling frequency), and the center-of-mass diffusion constant $D$ is
inversely proportional to the number of polymer segments (i.e., the length of
the polymer)\cite{de1979scaling}. Hence $\Delta L \propto 1 / \sqrt{L}$, where
$L$ is the length of the DNA molecule. For experimental setups identical to
ours, the choice of $k$ will vary based on the DNA length to be aligned
according to:
\[
k = k_{\text{ref}} \sqrt{\frac{L_\text{ref}}{L}}
\]
where $k_\text{ref} = 2$, as presented here, and $L_\text{ref}=24\mu m$, the length of the molecules presented above.

\noindent %

\subsubsection*{3. Shortest path finding}

Now every path between the peripheral nodes in our
networks represents a potential feature which satisfies the continuity
and completeness constraints above, so our task is to find the ``best''
such feature. Essentially, we are searching over all paths between the
peripheral nodes which do not move more than $k$ pixels horizontally
between any adjacent rows. Our cost function is the sum of pixel
intensities over a path, and we seek to minimize this over the space of these
acceptable paths. Thus, intuitively, the best bright feature corresponds to the
shortest path in $K_{B}$, and the best dark feature corresponds to
the shortest path in $K_{D}$. So, our task becomes a shortest path
finding problem.

Since our networks are directed and acyclic, the shortest path can
be computed by a standard dynamic programming algorithm based on topological 
sorting which grows linearly with the sum of the number of edges and
nodes (Fig. \ref{alg:WPAlign}, line 19) \cite{leiserson2001introduction}.
Note that, in our graph, this sum grows bilinearly
with the number of time frames and the horizontal width of our input
kymographs. And since the number of time frames does not
change, in practice, the sum of nodes and edges in our graph grows 
linearly with the kymograph width.
Thus the computational time of the shortest path finding algorithm 
also grows linearly with the kymograph width.

By this process we obtain two paths, one in $K_{B}$ corresponding to the best 
bright feature in our original kymograph, and one
in $K_{D}$ corresponding to the best dark feature. The path with the shorter
length of these two is chosen as the best overall feature 
(see fig. \ref{fig:kDandkBandNetwork}). If this feature is distinct enough (i.e., its length is lower than some threshold; see Methods on calculating $K_D$ and $K_B$) then we proceed with alignment. Otherwise the path is rejected, and recursion is terminated. This can happen, for instance, in large dark `gaps' associated with low labeling density mapping approaches.

\subsection*{Single Feature Alignment}

Once the best feature $F(y)$ has been
detected by the three-step process above, it is aligned by setting $F(y)=\langle F(y)\rangle$,
where $\langle .\rangle$ denotes the mean (rounding to the nearest integer) for all rows $y$, and the pixels
to the left and right of $F(y)$ are stretched (or compressed) linearly. To determine
intensity values at non-integer positions during this linear stretching, we used 
cubic spline interpolation (see Fig. \ref{fig:alignmentAndRecursion}
and Fig. \ref{alg:WPAlign}, line 23). Note that by setting $F(y)=\langle F(y)\rangle$,
we ensure that the feature is aligned over its position at thermodynamic
equilibrium.

\subsection*{Full Kymograph Alignment}

\begin{figure}
\begin{centering}
\includegraphics[width=0.8\textwidth]{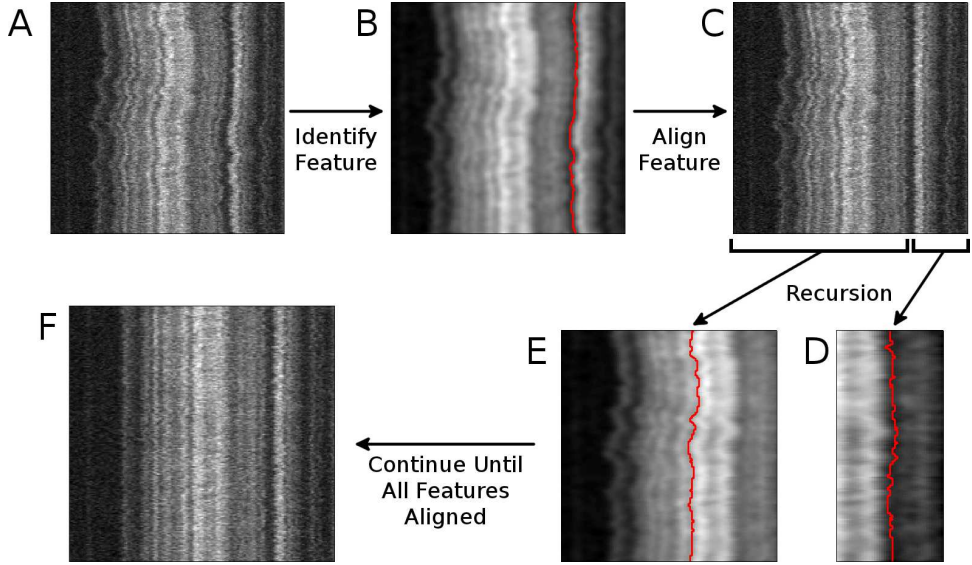}
\par\end{centering}

\caption{{\bf Feature detection and recursion in WPAlign.}
(A) A raw, unaligned kymograph is given as input (see Fig. \ref{fig:problemDefinition}).
(B) The ``best" feature is identified from the input kymograph.
(C) The feature identified in (B) is aligned via linear
interpolation.
(D, E) The feature identification process is called recursively on the regions
to the right (D) and to the left (E) of the newly-aligned feature.
(F) This process is continued until all features have been aligned.
\label{fig:alignmentAndRecursion}}
\end{figure}

\begin{figure}
\begin{center}
\includegraphics[scale=1]{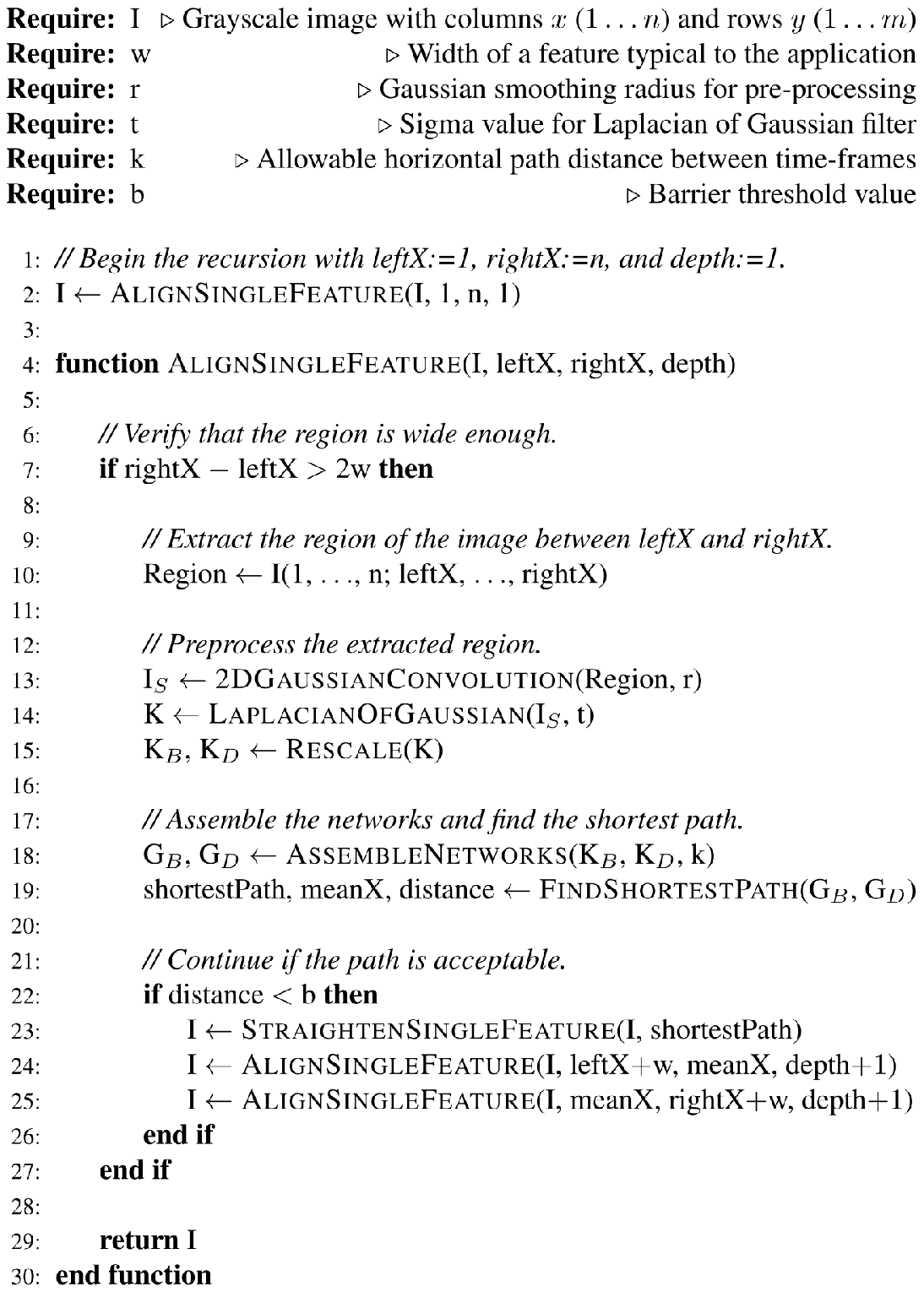}
\end{center}
\caption{{\bf WPAlign pseudocode.} Note that all substeps shown here are available in a variety of standard toolboxes, open-source and otherwise. For our particular implementation, all code was written in Matlab with dependencies in the Image Processing Toolbox and the Bioinformatics Toolbox (which implements the graph data structure and shortest path finding functionality).\label{alg:WPAlign}}
\end{figure}

Using the scheme above, the best feature in the input image has been both (1) \emph{detected} and
(2) \emph{aligned}. But we wish to detect and straighten 
all of the features. Thus to continue the process,
the image is split vertically along the newly aligned feature (see Fig.
\ref{fig:alignmentAndRecursion}c, d, and e), and $w$ columns (where $w$ is set to half 
the width of a typical feature) are removed from each on the side
adjacent to the split. Then the (1) {\it single feature detection} 
and (2) {\it single feature alignment} routines are called again on each of these two 
smaller images if they are
wider than typical a feature (i.e., $2w$). In our current application,
features are typically $\sim{10}$ pixels wide, so we use $w=5$ pixels, though this
parameter can easily be changed depending on the application.
(see Fig. \ref{fig:alignmentAndRecursion}, and Fig. \ref{alg:WPAlign}, lines 24--25).
The algorithm then terminates when the width of the image is less than the width of a
typical feature, $2w$.

It is worth noting that, since we remove columns before calling the process again,
the algorithm is guaranteed to converge. If we did not perform this step, 
the same feature could conceivably be re-straightened perpetually.

\noindent %

\section*{Results}

\begin{figure}
\noindent \begin{centering}
\includegraphics[width=0.8\textwidth]{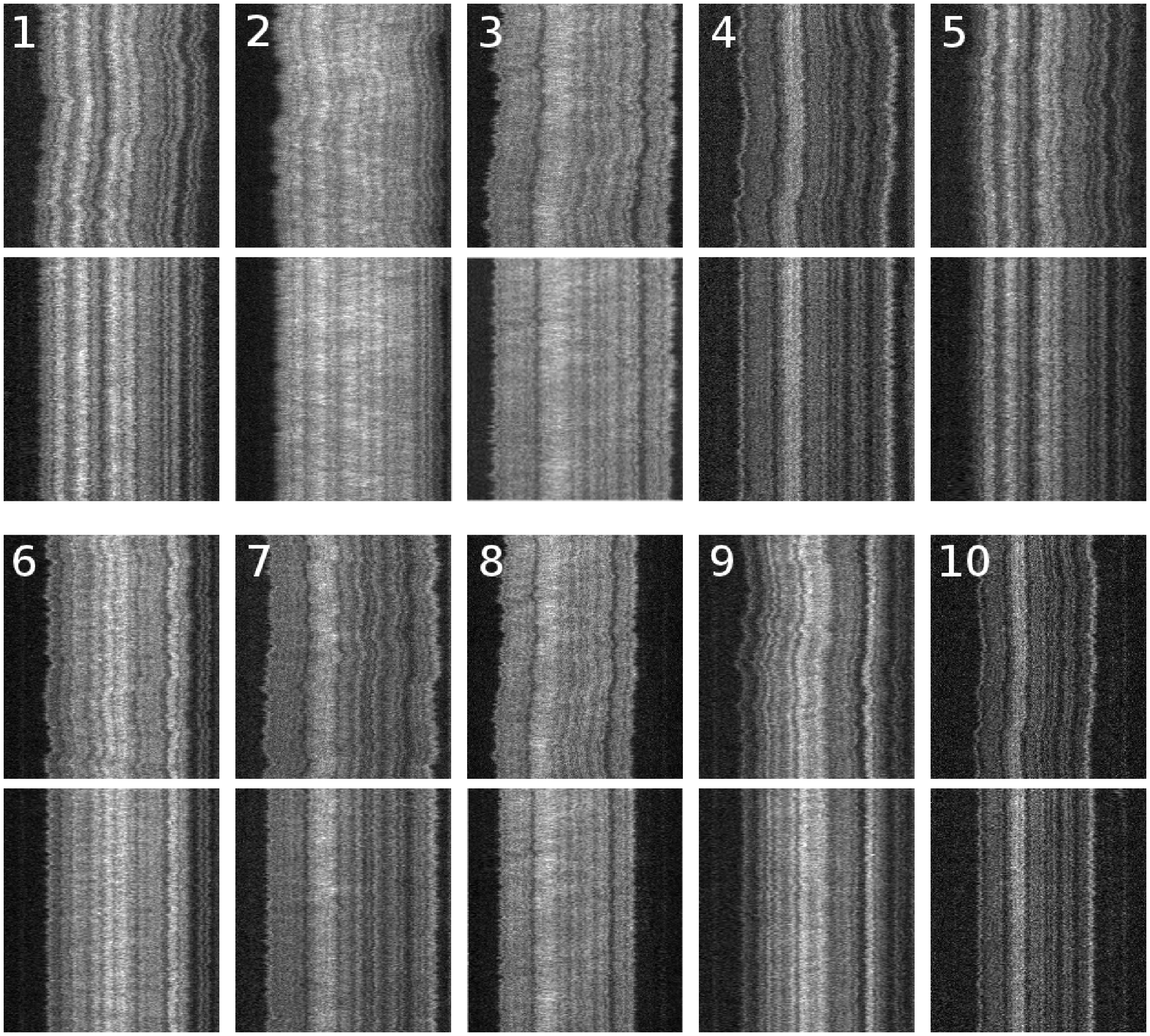}
\par\end{centering}
\noindent \centering{}\caption{{\bf Typical T4GT7 denaturation mapping kymographs, raw and
aligned via WPAlign.} The numbered kymographs represent the raw data, and the
aligned versions are displayed directly below each.\label{fig:AlignedKymographExamples}}
\end{figure}

In practice, WPAlign produces alignments which are visually
appealing (see Fig. \ref{fig:AlignedKymographExamples}). To quantify
the quality of these alignments, we use two quality criteria: \textit{time-trace
noise reduction} and \textit{information content} along with empirical computational costs.
We compare WPAlign to the method used in \cite{reisner2010single} (see
S1 for a description) with respect to these measures.

\noindent %

\subsubsection*{Time Trace Noise Reduction}

The ultimate goal of presenting the data in a kymograph format is to produce a 1-dimensional
intensity profile that is ``typical'' and reproducible for the DNA being analyzed.
This intensity profile is then compared to existing databases for genomic analysis. To produce this intensity profile,
$\langle I\rangle$, we simply take the time-average of an aligned
kymograph
\begin{equation}
\langle I(x,y)\rangle_y=\frac{1}{m}\sum_{y=1}^mI(x,y),
\end{equation}
where $m$ is the number of rows (time frames) in the kymograph. The reduction in noise can be quantified by the column-wise variance,
given by
\begin{equation}
\sigma^{2}(x)=\frac{1}{m}\sum_{y=1}^{m}\left[I(x,y)-\langle I(x,y)\rangle_y\right]^{2}.
\end{equation}

\begin{figure}
\noindent \begin{centering}
\includegraphics[width=0.7\textwidth]{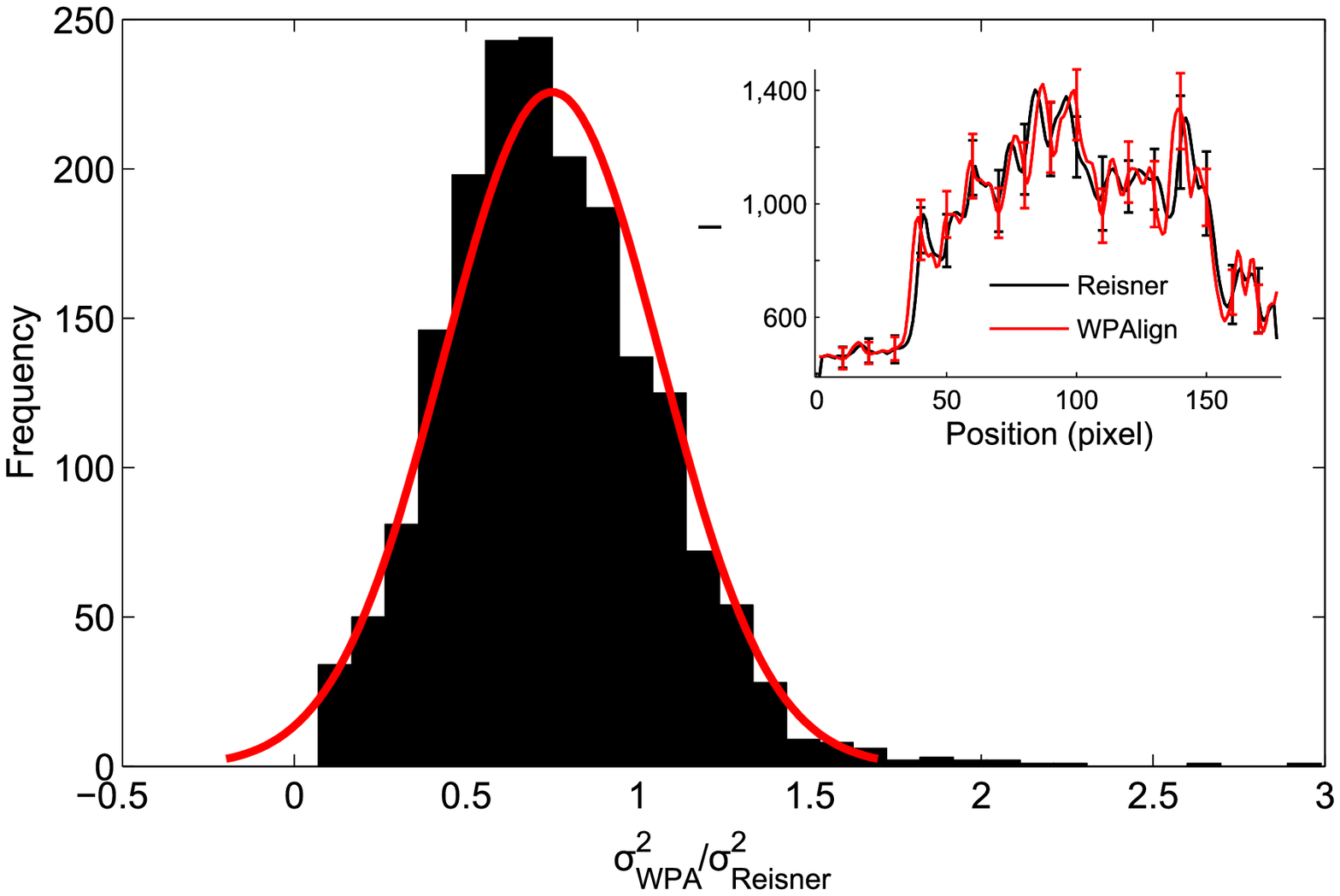}
\par\end{centering}

\caption{{\bf Kymograph noise comparison of WPAlign and the Reisner approach.} Column-wise
variances were calculated for aligned barcodes using WPAlign ($\sigma_{W}^{2}(x)$),
and the Reisner approach ($\sigma_{R}^{2}(x)$). Here we show the distribution
of $\sigma_{W}^{2}/\sigma_{R}^{2}$.
A Gaussian distribution was fit to this data, resulting in a mean
of $\mu=0.75$ (with standard deviation $\sigma = 0.32$). Thus in this particular case, WPAlign reduces
variance by 25\% compared to the Reisner method. Overall, the fraction of columns such that
$\sigma_{W}^{2} < \sigma_{R}^{2}$ was 80\%. (Inset) An example
time trace of barcode 6 (see Fig. \ref{fig:AlignedKymographExamples}) aligned via the WPAlign and Reisner
methods. Error bars represent one standard deviation 
(i.e., $\sigma_W$ and $\sigma_R$).\label{fig:NoiseComparison}}
\end{figure}

To compare WPAlign and the Reisner method\cite{reisner2010single}, these variances were calculated
for the 10 representative kymographs in Fig. \ref{fig:AlignedKymographExamples}, see S1 for experimental details on how these kymographs were obtained.
For every column in each of the kymographs, we obtained values $\sigma_{W}^{2}(x)$
and $\sigma_{R}^{2}(x)$, corresponding to the variances
resulting from alignment by the respective algorithms. Then a distribution
of values $\sigma_{W}^{2}(x)/\sigma_{R}^{2}(x)$
was calculated (see Fig. \ref{fig:NoiseComparison}). This distribution
is normal with mean $\mu=0.75$, indicating that, for a typical kymograph
column, WPAlign reduced variance by 25\% when compared to the Reisner
method. Thus WPAlign, besides reducing computational costs,
reduces kymograph noise and yields an improvement
over the Reisner approach.

\begin{figure}
\noindent \begin{centering}
\includegraphics[width=0.6\textwidth]{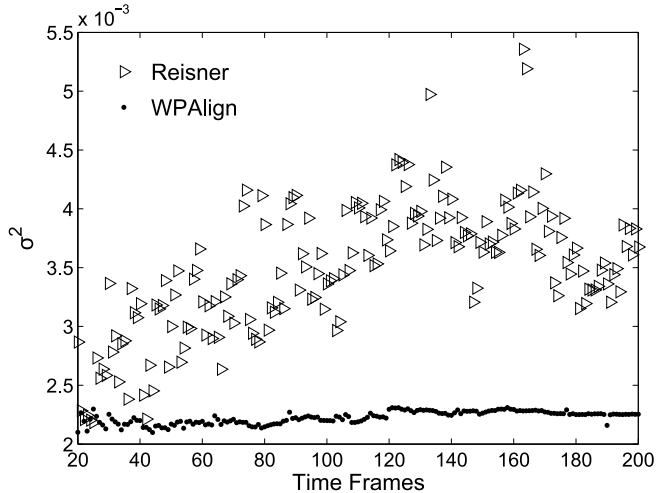}
\par\end{centering}

\caption{{\bf Effect of time axis length on aligned kymograph noise.}
Kymographs with time axes varying from 20 to 200 time frames  in length were
aligned by both methods. The resulting column-wise variances $\sigma^2_R(x)$ and $\sigma^2_W(x)$ were calculated
as in Fig. \ref{fig:NoiseComparison}. Here we plot, for each kymograph,
the mean of these column variances, i.e.,
$\langle\sigma^2_R(x)\rangle$ and
$\langle\sigma^2_W(x)\rangle$, showing that kymograph noise increases with time axis length for the Reisner method
but remains constant for WPAlign.
\label{fig:NoiseTimeDep}}
\end{figure}

In addition, we examined how aligned kymograph noise is affected by the
length of the time axis (see Fig. \ref{fig:NoiseTimeDep}).
To do this, we calculated $\langle\sigma^2(x)\rangle$, the mean of the column-wise variances given above, given by
\begin{equation}
\langle\sigma^{2}(x)\rangle=\frac{1}{n}\sum_{x=1}^{n}\left[\frac{1}{m}\sum_{y=1}^{m}\left[I(x,y)-\langle I(x,y)\rangle_y\right]^{2}\right],
\end{equation}
for kymographs with time axes ranging from 20 to 200 time frames.
From the results we can see that the noise from kymographs aligned via
WPAlign is independent of the number of time frames chosen during
filming. However, kymographs aligned by the Reisner approach
seem to undergo an increase in noise as the number of frames
increases, until plateauing at roughly 120 frames.

We present a possible explanation for this result:
as the number of time frames increases, the underlying DNA molecule 
is allowed more time to undergo conformational
changes and random diffusive processes, 
rendering the first and last frames
increasingly dissimilar. Thus any template frame chosen by
the Reisner method will be increasingly dissimilar from the frames farthest from it
in the kymograph. This renders the local stretching factor
optimization more prone to becoming stuck in local
minima for these ``distantly related" frames, introducing noise
in the final alignment. WPAlign, on the other hand, avoids this 
problem as it does not rely on the choice of a single template frame.

\noindent %

\noindent %

\subsubsection*{Information Score}

Before alignment, bright and dark features can stray into adjacent
kymograph columns as the DNA molecule undergoes horizontal diffusion.
Thus thin features are obscured in the time average, effectively broadening
peaks and valleys. As alignments improve, however, features occupy
less horizontal space and become more apparent in the final
time-trace. Since these features are essential in later performing
statistical comparisons with other data, increasing contrast between
neighboring features (i.e., increasing feature sharpness) 
essentially increases the information contained in the time trace.

To  quantify this information content of the barcodes we present a new scheme
based on the self-information of a random variable\cite{informationTheory}, see 
Methods for a comprehensive description.
Here in brief, the information score, $\text{IS}$, of a kymograph is given by
\begin{equation}
\text{IS}=\sum_{k}-\log\left[\frac{1}{\sqrt{2\pi\log(\sigma^{2}+\chi)}}\exp\left\{
    -\frac{\log(\lvert\Delta I_{k}{}\rvert)^{2}}{2\log(\sigma^{2}+\chi)}\right\}
\right]
\end{equation}
 where $\sigma^{2}$ represents the noise of the underlying kymograph, $\chi=1$ is
 a regularization parameter ensuring that the score remains real-valued for all
 noise levels,
and the $\Delta I$'s represent ``robust" intensity differences (see S1)
between neighboring peaks and valleys. Thus information increases as the
difference between neighboring peaks and valleys grows and as the noise of the
underlying kymograph decreases. That is, time-traces with sharper, more
well-defined local extrema will contain more information than time traces from
noisy kymographs with broad extrema.

\begin{figure}
\noindent \begin{centering}
\includegraphics[width=0.9\textwidth, height=0.18\textwidth]{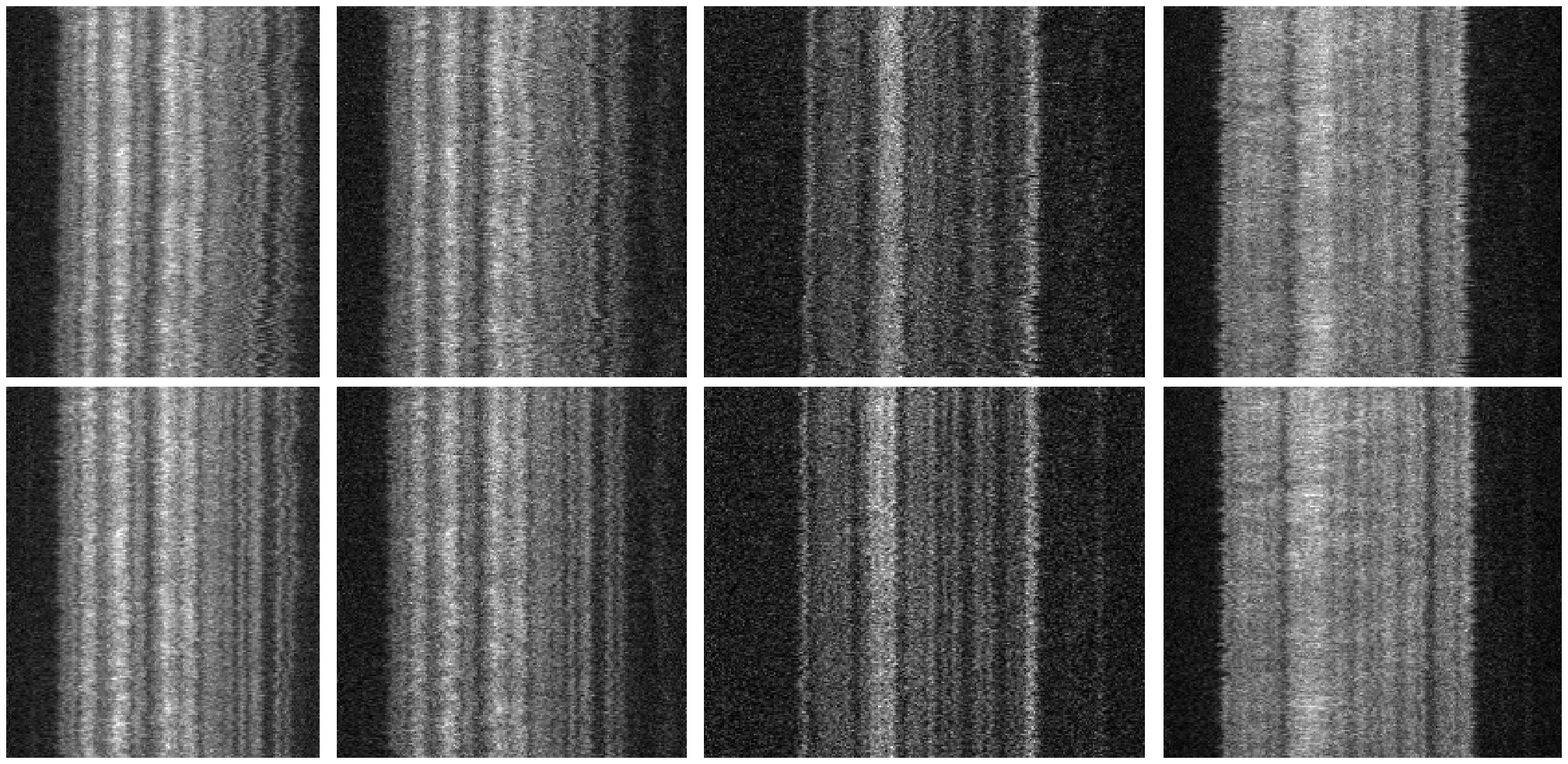}\\
\vspace{1em}
\includegraphics[width=\textwidth, trim=2.5cm 0cm 2.5cm 0cm]{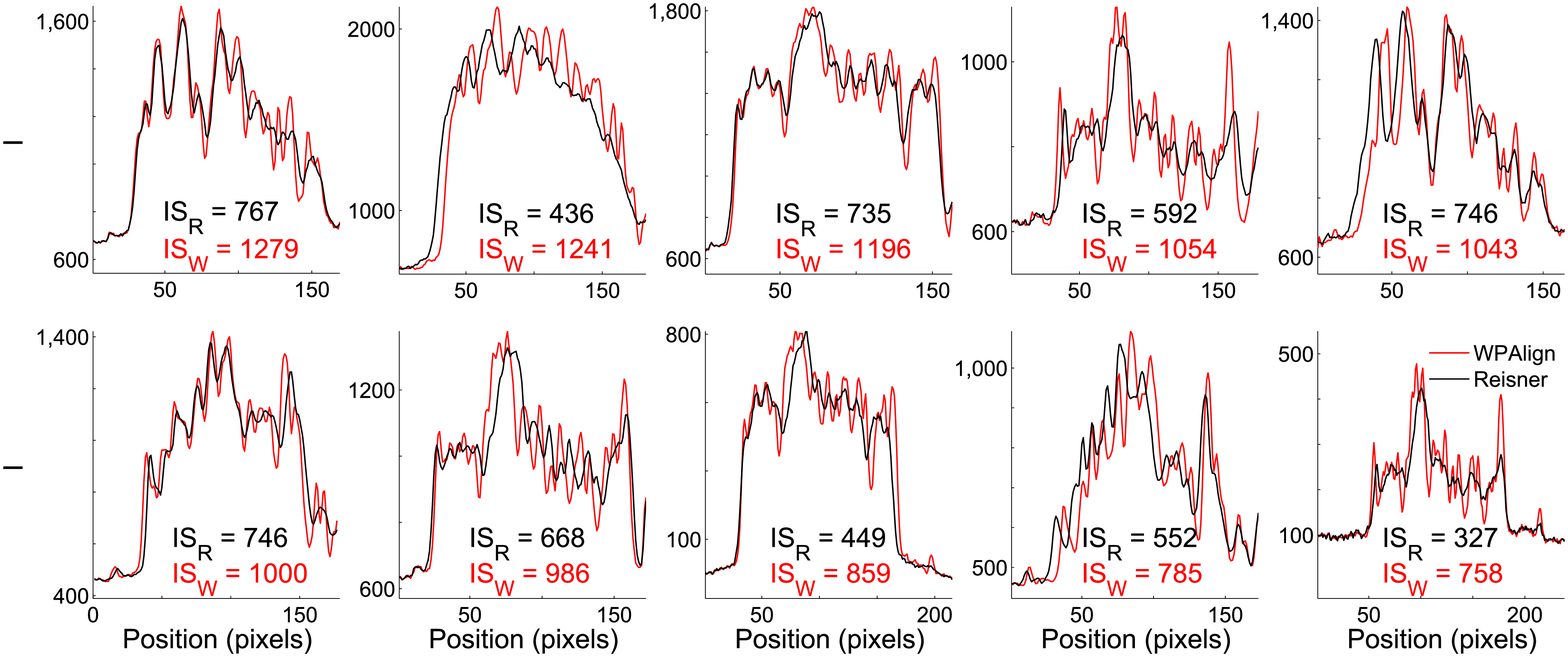}
\par\end{centering}

\caption{{\bf Quality comparison of barcodes aligned via the Reisner approach and WPAlign.}
(Top) Representative kymographs aligned via the Reisner approach (above) and WPAlign (below).
(Bottom) Average intensity traces from kymographs in Fig. \ref{fig:AlignedKymographExamples} aligned via WPAlign (red) and the Reisner approach (black). The information score is displayed below each trace for both methods, quantifying the contrast between neighboring features. Notably, $\langle (\text{IS}_{W}-\text{IS}_{R})/\text{IS}_{R}\rangle=0.78$, indicating that WPAlign on average produced time-traces with slightly more information than the Reisner method over our sample set. Note that the plots are in the same order as the corresponding kymographs in Fig. \ref{fig:AlignedKymographExamples} and are ordered by decreasing $\text{IS}_W$. \label{fig:InformationContent}}
\end{figure}

Time traces and their corresponding information scores were calculated
for the representative kymographs shown in Fig. \ref{fig:AlignedKymographExamples}
after undergoing alignment by both WPAlign and the Reisner method
(see Fig. \ref{fig:InformationContent}). In general, WPAlign
produces visually sharper and more well-defined time-traces. Furthermore,
the information content is greater for WPAlign in all examples considered.
In fact, the average information gain from using WPAlign over the
Reisner method on our dataset is roughly 78\% (i.e., $\langle (\text{IS}_{W}-\text{IS}_{R})/\text{IS}_{R}\rangle=0.78$).

Furthermore, independent of our kymograph alignment technique, this information
score can serve as an objective and easily interpretable barcode quality measurement
by which barcodes can be compared, providing a basis for experimental optimization.
For example, expected barcodes can be calculated from theory for a number
of experimental conditions\cite{reisner2010single}, and then experiments can be performed only for
barcodes expected to yield the highest information content, saving valuable
time and resources for experimentalists. Furthermore, if theory is not available for a
particular optical mapping application, the information score can serve as a
quality criterion for experimental barcodes by which experimental conditions
can be rigorously optimized.

\noindent %

\subsubsection*{Computational Time}

\begin{figure}
\noindent \begin{centering}
\includegraphics[width=0.55\textwidth]{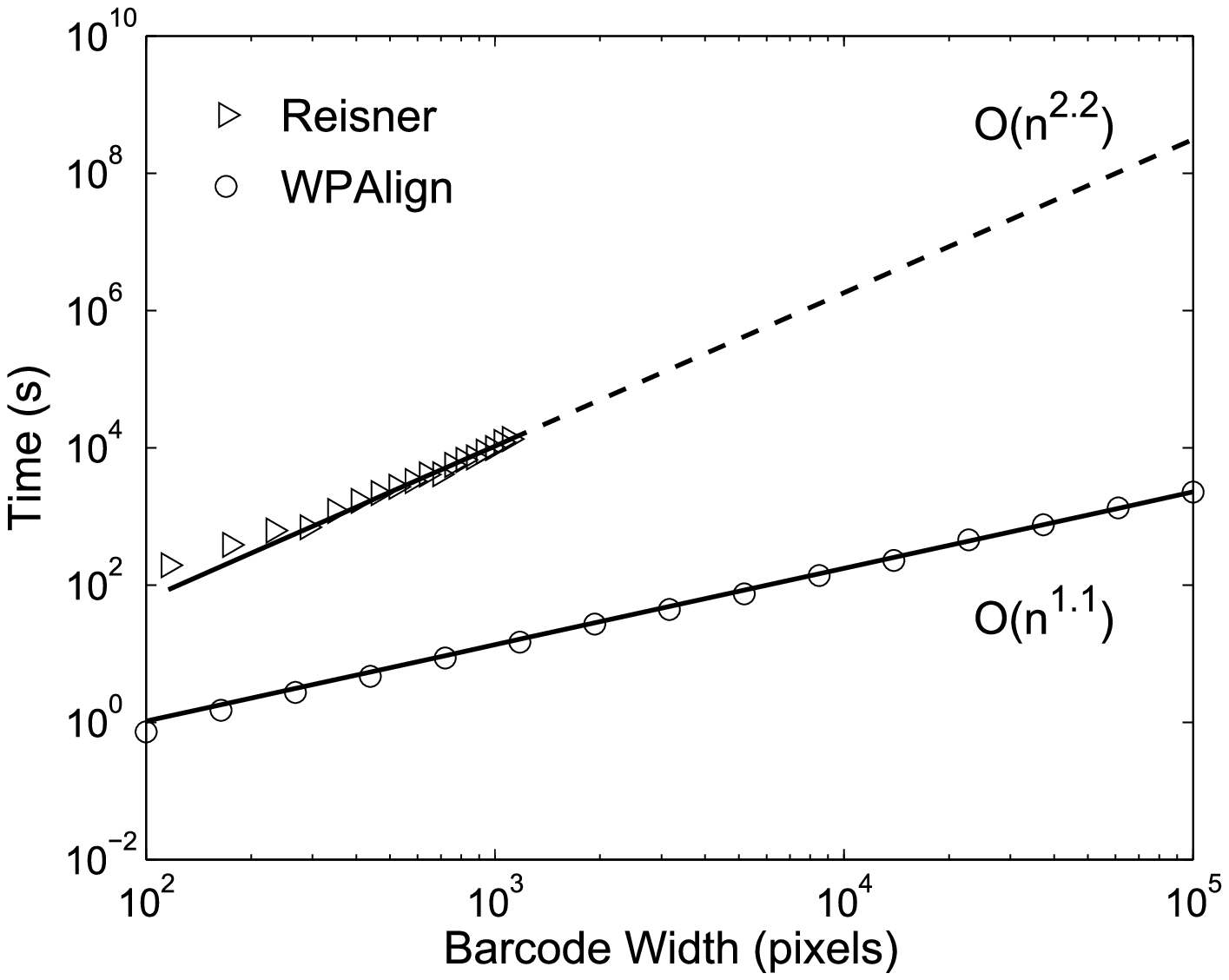}
\par\end{centering}

\caption{{\bf Comparison of time scaling for WPAlign and the Reisner approach.} 
Empirical time scaling results for both techniques on identical kymographs
ranging between $\sim{10^2}$ and $\sim{10^5}$ pixels in width, where
$\sim{10^5}$ pixels is roughly the length of an intact human genome
at current resolutions.
Power law curves of the form $ax^b$ were fit to these data (solid lines), yielding $b=1.1$ for 
WPAlign, and $b=2.2$ for the Reisner method. Thus WPAlign
exhibits approximately $O(n)$ time scaling with barcode width, while
the Reisner method scales approximately with $O(n^2)$.
Scaling data beyond $\sim{10^3}$ pixels was projected for the Reisner approach
(dashed line), as alignment times became prohibitive.
Simulated kymographs were generated by concatenating experimental T4GT7 kymographs
from above. \label{fig:Scaling}}
\end{figure}

Empirically, WPAlign exhibits linear scaling with barcode length, $n$ 
(see Fig. \ref{fig:Scaling}).
This is somewhat intuitive, as the bottleneck of the approach is
the shortest path finding algorithm, and this runs in linear time
due to the directed and acyclic qualities of our networks. The Reisner
approach, on the other hand, scales with $O(n^{2})$, rendering it
impractical for bacterial barcodes on the order of 1 Mbp or larger 
(see Fig. \ref{fig:Scaling}).

Perhaps most importantly, WPAlign was able to successfully align a
simulated barcode with length on the order of a full human genome
in only 40 minutes. The Reisner approach would require over
$10^{8}$ seconds, or roughly 3 years, to perform this task
on an identical computer.

\noindent %

\section*{Discussion}

In this paper we present a new DNA barcode kymograph alignment algorithm which outperforms an existing alternative \cite{reisner2010single} in computational speed, and for the particular data presented here, also slightly improves on noise reduction properties and information content of the time-averaged barcodes. Indeed, the orders-of-magnitude improvement in computational speed could open the door for high throughput kymograph alignment at the human-genome scale as well as constituting an important step in data analysis for a number of nanofluidic optical mapping techniques, including denaturation mapping\cite{reisner2010single}, fluorocoding\cite{fluorocode}, competitive binding\cite{competitiveBinding, nilsson2014competitive}, and enzymatic nicking\cite{enzNicking1,enzNicking2}. By providing a rapid framework for this data analysis, WPAlign can help bring the many applications of optical mapping closer to application, including bacterial strain and species identification, detection of large-scale genomic structural variation, and scaffold building for third generation \textit{de novo} sequencing techniques. Furthermore, the algorithm is easy to implement, as the various substeps involved are available in most standard numerical packages.

Moreover, our feature detection method, suitably modified, may find
application in other domains of biological image analysis, such as automated organism 
tracking\cite{tsibidis2007nemo,huang2008automated}, or automated study of cellular transport
along axons (\`{a} la \cite{chetta2011novel,EJN:EJN12263,zhou2001direct,Miller01062004}), defects in which have been implicated in a number of
neurodegenerative diseases, including Alzheimer's Disease, Parkinson's
Disease, and Amytrophic Lateral Sclerosis\cite{chevalier2006axonal,haghnia2007dynactin}.

Finally our information score, motivated as a way to compare the quality of
alignments produced by different methods, may find widespread application
among the optical mapping community as an easily calculable and interpretable
barcode quality criterion by which to optimize experimental conditions.

\section*{Acknowledgments}
We thank Karl {\AA}str\"{o}m for helpful discussions on image analysis.

\bibliography{writeup_final_arxiv}



\section*{Supporting Information}

\subsection*{Kymograph Generation}

As raw data, we are given movies of fluorescently stained DNA molecules confined in nanochannels. To generate kymographs from these movies, we perform the following procedure. First, we create a single representative frame via a simple time average of the entire movie. This frame is then rotated so that the channels are horizontal, and individual molecules are detected via image segmentation. For a given molecule, a 1D intensity trace is calculated for each movie frame by averaging over a 3-pixel vertical window, and these 1D traces are then stacked to form a kymograph.

\subsection*{The Reisner Method}

Here we explain our implementation of the Reisner alignment method, as described in \cite{reisner2010single}. As usual, we begin with a raw, unaligned kymograph produced as detailed above. To begin the alignment, a template row, $T$, is chosen (typically picked near the middle of the kymograph), and then each non-template row, $A_i$, is individually and independently aligned to the template row in the following way.

First, all the $A_i$ are translated linearly so that their centers of mass are aligned to $T$'s. Then the next step is to ``smooth out" the local longitudinal fluctuations. This is done by dividing the $A_i$ into a series of uniform length pieces and applying a set of dilation/contraction factors to them.

In practice, a piecewise linear map $S_i$ is created, and the slopes of the individual linear components are defined by dilation factors $d_k$. Thus $S_i$ itself is a function of the $d_k$, for example $S_i(d_k)$. Now this map $S_i$ operates on the row $A_i$ we wish to align, and a new profile, $A_i^{'}(x_j,d_k)$ is created (where $x_j$ is the $j$th pixel in the profile). The parameters $d_k$ are chosen to minimize the least squared difference $\Delta$ between the row $A_i$ and the template row $T$:

\[
\Delta = \sum_{j=1}^{N} \left[  A_i^{'}(x_j,d_k) - T(x_j)  \right]^2
\]

Minimizing $\Delta$ with respect to the $d_k$ can be performed by any standard global optimization procedure; however it must be noted that the process is very susceptible to local minima. Thus in our implementation we employ simulated annealing; in this process  we limit the number of $\Delta$ evaluations to a number which increases linearly with the number of dilation factors $d_k$.

Once the dilation factors have been chosen for all rows, they are normalized so that the average dilation $d_k$ across all rows is one (i.e., $\langle d_k(i) \rangle = 1$). This is done to better approximate the true equilibrium conformation of the DNA and help minimize the effect of template frame choice. Now the $d_k(i)$ are applied to all rows $A_i$ to obtain the final aligned kymograph.

\subsection*{Laplacian of Gaussian Filter}

The Laplacian of Gaussian filter is a standard image processing technique which is useful for ``blob" detection. It uses the sum of second derivatives in the image to emphasize blobs of size roughly given by the variance of the Gaussian kernel. In this way, we emphasize not so much edges as ridges and valleys in the data which will be easier to detect in our feature detection step.

To perform the Laplacian of Gaussian, first we convolve $I$ with a Gaussian kernel
\begin{equation}
g(x,y,t)=\frac{1}{2\pi t}\ \mbox{exp}\left\{ -\frac{x^{2}+y^{2}}{2t}\right\}
\end{equation}
to give a scale space representation $L(x,y;t)=g(x,y,t)\star I(x,y)$.
Then the Laplacian operator $\nabla^{2}L=L_{xx}+L_{yy}$ is computed,
resulting in strong positive responses for dark regions of extent
$\sqrt{2t}$ and strong negative responses for bright regions of similar
extent \cite{LoG}. For our data, we have found $t=10$ pixels to
be adequate.

The size of the applied filter is set to be 10 pixels in the horizontal direction and only 3 pixels in the vertical direction, rendering the process close to one dimensional but with a small vertical component. This is done because features are expected to fluctuate horizontally between vertical time frames.

\subsection*{Calculating $K_{D}$ and $K_{B}$}

Given a Laplacian response $K(x,y)$, linearly scaled such that each value falls in
the range $[-1, 1]$, we calculate $K_{D}$ and $K_{B}$ using
\begin{equation}
K_{D}(x,y)=\begin{cases}
1-K(x,y) & \mbox{for }K(x,y)>0\\
B & \mbox{for }K(x,y)\leq0\end{cases}
\end{equation}

\begin{equation}
K_{B}(x,y)=\begin{cases}
1+K(x,y) & \mbox{for }K(x,y)<0\\
B & \mbox{for }K(x,y)\geq0\end{cases},
\end{equation}
where $B\gg1$ is a large constant which creates {}``barrier''
pixels through which paths will not traverse. In this way, we prevent
the feature detection algorithm from {}``jumping'' between adjacent
features. Then in $K_{D}$, small values represent dark regions, and
in $K_{B}$ small values represent bright regions.

\subsection*{Information Score}

Here we introduce an information score associated with a DNA barcode,
given as a 2D intensity profile, $I$.
This score was chosen to quantify the amount, and sharpness, of
{}``robust'' peaks and valleys in the barcode.

We define these ``robust" extrema as those which differ from 
neighboring pixels by an amount greater than a threshold value, $I_{\rm th}$, typically chosen to be equal to the background noise level. In order to
quantify this background noise, we assume that $I$ has already been aligned.
By the nature of the alignment, each column
in $I$ represents a single intensity value obscured by the addition
of noise due to the photophysics of the dyes, noise in the imaging system and thermal
fluctuations of the confined DNA molecules. Thus we create a new image whose pixel in the $y$th
row, $x$th column, is given by
\begin{equation}
I'(x,y)=I(x,y)-\langle I(x,y)\rangle_{y},
\end{equation}
where $I(x,y)$ is simply the intensity of this pixel in the aligned
kymograph, and $\langle I(x,y)\rangle_{y}$ is the mean intensity of the
$x$th column of the kymograph. Then, assuming perfect alignment,
$I'$ is an image with intensities
attributable only to our kymograph's background noise, so our background
variance $\sigma^{2}$ is given by the intensity variance of $I'$,
or
\begin{equation}
\sigma^{2}=\langle(I'-\langle I'\rangle)^{2}\rangle.
\end{equation}
Now we calculate the time average of $I$, denoted $\langle{I(x,y)}\rangle_y$, given by
\begin{equation}
\langle{I(x,y)}\rangle_y = \frac{1}{m}\sum_{y=1}^mI(x,y)
\end{equation}
and we locate the ``robust" peaks and valleys by treating
$\langle{I(x,y)}\rangle_y$ as an energy landscape and employing the method of
Azbel, developed for a simplified description of the random melting of a one
dimensional two-component Ising model \cite{azbel1973random,azbel1980dna} (see
below).  Thus we obtain only local extrema which are ``robust," i.e., regions
which to the left and right are surround by an ``energy barrier" larger than
the threshold $I_{\rm th}$ (typically chosen equal to the average background noise,
i.e., $\sigma$).

\begin{figure}
\noindent \begin{centering}
\includegraphics[width=0.4\textwidth]{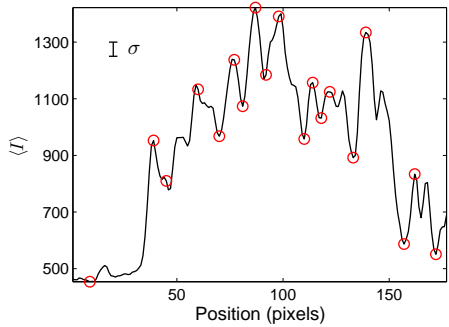}
\par\end{centering}

\caption{{\bf Illustration of ``robust" extrema detected by Azbel's method.} The aligned time-average of kymograph 6 (from Figure \ref{fig:AlignedKymographExamples}) was used, with $\sigma=50$ intensity units for illustrative purposes.\label{fig:Azbel}}
\end{figure}

Let us now, for completeness, briefly review the method by Azbel
\cite{azbel1973random,azbel1980dna} for finding robust local maxima and
minima. The method scales linearly with the barcode length and uses the fact
that robust local maxima and minima are alternating, i.e. a local maximum must
be followed by a local minimum and vice versa. Since the barcode's initial
pixels in general represent background noise, in our version of the Azbel
approach we utilize that the first local extremum must be a minimum. The
algorithm now proceeds as follows: we step through the pixels, $x$,
consecutively. For each pixel we calculate the intensity difference 
$C(x_s,x)= \langle{I}\rangle_x-\langle{I}\rangle_{x_s}$ where $x_s$ is a start pixel as
defined below. At the start of the algorithm we set $x_s=1$ and calculate
$C(x_s,x)$ for increasing $x$ as long as the criteria
$C(x_s,x)<I_{\rm th}$ is fulfilled (implemented through a while-loop). Once
this criterion is violated we know that the pixel $x_s$ must be a robust local
minimum (surrounded to the right by a barrier larger than $I_{\rm th}$), so
this pixel number is stored and the while-loop terminated. If, within the
while-loop above, the intensity difference decreases, i.e. we find that
$C(x_s,x) <0 $, we must have a new local minima and we shift $x_s$ to the
present pixel position, $x_s\rightarrow x$, with a corresponding reset of
$C(x_s,x)$. Once a robust local minima has been identified according the the
scheme above, we proceed in an identical fashion to identify a subsequent
robust local maximum, starting at pixel $x$. To this purpose we invoke the
condition $C(x_s,x)>-I_{\rm th}$ (and $C(x_s,x)>0 $ as a requirement for
shifting $x_s$) within a new while-loop. The procedure above is repeated until
all $n$ pixels representing $\langle{I(x,y)}\rangle_x$ has been exhausted. In
Fig. \ref{fig:Azbel} we display the local extrema found using the approach above.

\noindent %

Finally our information score is that of \textit{self information}\cite{informationTheory},
and we define the information content $\text{IS}$ of a DNA barcode by
\begin{equation}
\text{IS}=\sum_{k}-\log\left[\frac{1}{\sqrt{2\pi\log(\sigma^{2}+\chi)}}\exp\left\{ -\frac{\log(\lvert\Delta I_{k}{}\rvert)^{2}}{2\log(\sigma^{2}+\chi)}\right\} \right],\label{eq:J}
\end{equation}
where the $\Delta I_k$s are the intensity differences between neighboring peaks and valleys identified above, and $k$ denotes the numbering of $\Delta I$s found. Also, $\chi=1$ is a regularization parameter which ensures that IS is real-valued for all noise levels.

To justify the use of the logarithm of $\Delta I(k)$ rather than
$I(k)$ itself in Eq. (\ref{eq:J}) we note that in a simplistic approach
to DNA melting we have that the probability for a DNA basepair to
be open is related to the Boltzmann weight $P(i)\propto\exp(-\beta\Delta E(i))$,
where $\Delta E(i)$ is the energy difference between a basepair being
open and closed. Therefore, by using ${\rm log}[I(k)]$ our information
score utilizes free energy differences.

\subsection*{Experimental Procedure}

To generate the optical DNA mappings shown in the main text, we employed the following 
experimental protocol. T4GT7 DNA (supplied by Nippon Gene, Japan through Wako Chemicals GmbH, Neuss, Germany) was mixed at 
a ratio of 1 dye molecule per 6 base pairs with YOYO1\textregistered \,\,Iodide 
(LifeTechnologies\textregistered, USA) and kept at 50$^\circ$C for 2 hours to ensure homogeneity of staining. 
Melting experiments were performed in a buffer consisting of 10mM NaCl in 0.05 x TBE 
(1 x TBE is 89mM Tris, 89mM Boric acid and 2mM EDTA). Beta-mercaptoethanol was 
added to a final concentration of 2\% and formamide to a final concentration of 50\%.

Using compressed nitrogen the buffer carrying the DNA molecules was forced into 
nanochannels with cross section 100 nm by 150 nm etched in fused silica. Once in the nanochannels the 
molecules were imaged in a Nikon TE2000 microscope (Nikon, Tokyo, Japan) using a 
high-pressure mercury lamp for excitation, a FITC filter cube to pick out the 
fluorescence from the stained DNA and an Andor Ixon DU 897 EMCCD (Andor Technology, 
Belfast, Ireland) camera for image acquisition.

In order to form the melt maps the nanochannel device was brought into contact with 
an aluminium block, heated to 31.5$^\circ$C at which temperature the molecules are partially 
denatured. We utilized the time dependence of the barcode formation process to create kymographs of different quality and different information contents. Molecules were imaged over several distinct 5-minute periods resulting in the 10
kymographs in Figure \ref{fig:AlignedKymographExamples}, representing a collection 
of 4 molecules. Kymographs generated from the same molecules were: (1, 2, 5), (3, 8), (7, 4, 10), (6, 9).

\end{document}